# Epitaxial Growth of Large Area Single-Crystalline Few-Layer MoS$_2$ with Room Temperature Mobility of 192 cm$^2$V$^{-1}$s$^{-1}$

Lu Ma[1], Digbijoy N. Nath[2], Edwin W. Lee II[2], Choong Hee Lee[2], Aaron Arehart[2], Siddharth Rajan[2]*, Yiying Wu[1]*

[1]Department of Chemistry & Biochemistry, The Ohio State University, 100 West 18$^{th}$ Avenue, Columbus, Ohio 43210.

[2]Department of Electrical and Computer Engineering, The Ohio State University, Columbus, Ohio 43210.

ABSTRACT

We report on the vapor-solid growth of single crystalline few-layer $MoS_2$ films on (0001)-oriented sapphire with excellent structural and electrical properties over centimeter length scale. High-resolution X-ray diffraction scans indicated that the films had good out-of-plane ordering and epitaxial registry. A carrier density of ~2 x $10^{11}$ $cm^{-2}$ and a room temperature mobility of 192 $cm^2$/Vs were extracted from space-charge limited transport regime in the films. The electron mobility was found to exhibit in-plane anisotropy with a ratio of ~ 1.8. Theoretical estimates of the temperature-dependent electron mobility including optical phonon, acoustic deformation potential and remote ionized impurity scattering were found to satisfactorily match the measured data. The synthesis approach reported here demonstrates the feasibility of device quality few-layer $MoS_2$ films with excellent uniformity and high quality.

Recently, there has been a rapidly increasing interest on investigating layered 2-dimensional (2D) materials such as $MoS_2$, $WS_2$, $WSe_2$ *etc.* for their promise towards a variety of next-generation electrical[1,2] and optoelectronic[3,4] device applications including low cost, flexible[5] and transparent electronics.[6,7] From an epitaxial point of view, these materials circumvent limitations associated with lattice mismatch in heterostructure growth of conventional semiconductors and could enable growth of crystalline epi-layers highly lattice mismatched with substrate.[8] Among the 2D materials of the dichalcogenide ($MX_2$) family, field effect transistors (FETs) with high on/off ratio and high current densities were fabricated based on $MoS_2$,[1,2] $MoSe_2$,[9] and $WS_2$,[10] while both p-FET and n-FET based on $WSe_2$,[11,12] were also reported besides demonstration of $MoS_2$-based simple integrated circuits.[13]

Most of the $MoS_2$-based devices reported till date have been fabricated on flakes of $MoS_2$ mechanically exfoliated from geological samples. These exfoliated micro-flakes of $MoS_2$ have randomly distributed thickness and orientation, and are not viable for large-scale device integration. To achieve epitaxy of large-area $MoS_2$ films with uniformity and control, chemical vapor deposition (CVD) methods using various precursors such as $MoO_3$,[14-18] $MoO_2$,[19] $MoCl_5$,[20] $Mo$[21,22] and $(NH_4)_2MoS_4$[23,24] or physical vapor transport method[25] have been employed. However, these large-area films are mostly polycrystalline with small crystal grain sizes from tens of nanometers to several micrometers. In this report, we demonstrate large area vapor-solid-grown epitaxial $MoS_2$ thin films on sapphire with in-plane and out-of-plane ordering over centimeter length scales. The high quality of these films leads to record high carrier mobility (~200 $cm^2$/Vs) for synthetic films at room temperature and current density in excess of 0.15 mA/µm. This work demonstrates the feasibility of device quality few layer $MoS_2$ layers with sufficient uniformity and quality to enable a variety of device applications based on 2D layered materials.

The vapor-solid grown MoS$_2$ films are of significantly higher crystalline quality than previous attempts.[21] This can be attributed to three main factors - use of epitaxial substrates, control of grain nucleation, and high growth temperatures. We used an epitaxial template (sapphire, space group: R-3c) that shares some symmetry with the MoS$_2$ (space group: P6$_3$/mmc) structure, and is thermally stable up to relatively high temperature. It has been shown previously that under similar conditions, the use of an epitaxial template led to significantly better crystal quality.[16, 24, 26-28] In previous reports of CVD-grown MoS$_2$, while out of plane ordering was evident from X-ray diffraction (XRD) measurements in the (0001) direction, no in-plane ordering, such as that evidenced from off-axis geometry, was apparent. We attribute this to significant twist in the mosaic caused by the large number of independent grains.

In this work, the control of grain nucleation was achieved by the supersaturation of sulfur vapor. Single-crystal (0001)-orientated sapphire substrates were solvent cleaned and 5 nm of molybdenum was deposited by sputtering using AJA Orion RF/DC Sputter Deposition Tool. 8.0 mg of MoS$_2$ powder (purchased from Sigma Aldrich) was placed in a small quartz tube which was then put inside a larger quartz tube (inner diameter: 1 cm) along with the Mo-coated sapphire wafer. The larger quartz tube was pumped down by a mechanical pump, sealed and heated to 1100ºC for 4.5 hours and then cooled down to room temperature at a rate of 0.5 ºC/min.

Nucleation is the first step in the crystallization process. Nuclei formed in this step can continue to grow into crystalline domains. In order to get single-crystal MoS$_2$ films with good crystalline nature, the nuclei density should be minimized, which can be achieved by a controlled low supersaturation. In our vapor-solid synthesis, initially sulfur vapor reacts with the surface Mo metal and creates the seed crystal. Sulfur pressure needs to stay at a low value to reduce the nucleation density. It is difficult to control the amount of sulfur by using elemental sulfur itself because the super saturation is very high compared to the small amount of Mo metal on the substrate. We reduced the sulfur pressure by using $MoS_2$ powder as a sulfur source to get a low sulfur pressure during the synthesis (Figure 1). Thermogravimetric analysis (TGA) of $MoS_2$ shows that above 950ºC, sulfur can be gradually released and detected by mass spectrometer.[29-32] The decomposition of $MoS_2$ powder at high temperature can provide sulfur to sulfurize the Mo metal on sapphire substrate. The equilibrium between $MoS_2$ powder and sulfur vapor inside the quartz tube would provide a sulfur vapor pressure of 0.023Pa at 1100ºC.[30] According to the pressure-temperature phase diagram of Mo-S system,[33] this pressure is the lowest sulfur pressure at which pure $MoS_2$ phase can be produced. The total amount of sulfur that $MoS_2$ can release under vacuum for 4.5 hours is about 0.099%,[30] which is calculated to be the amount of sulfur that the Mo layer on substrate needed to get $MoS_2$ phase (calculation result in Supporting Information).

During the synthesis, the following reactions occur:

$$MoS_2 \text{ powder (surface)} \rightarrow MoS_x \ (x < 2) + \frac{2-x}{2} S_2 \qquad (1)$$

$$Mo \text{ (on sapphire)} + S_2 \rightarrow MoS_2 \text{ (on sapphire)} \qquad (2)$$

(0001)-orientated sapphire was chosen to be a good Van der Waals epitaxy substrate due to its atomically flat surface without dangling bonds on the surface. Both theoretical[34, 35] and experimental [36, 37] study show that the surface of (0001)-orientated sapphire is terminated by one Al layer because in this situation the dangling bonds on the surface are either completely filled or empty to form an auto-compensated neutral surface. Thus, the lattice matching condition has been relaxed dramatically.

The surface morphology of the samples was characterized by atomic force microscope (Veeco Instruments DI 3000). Crystalline nature was examined by a Bruker D8 High-Resolution Triple Axis X-Ray Diffractometer. Raman spectra were obtained by Renishaw spectrometer with a 10 mW laser at 514 nm.

Device fabrication started with standard lithography using an i-line stepper projection aligner followed by e-beam evaporation of Ti/Au/Ni metal stack for Ohmic contact. The devices were then mesa isolated using $BCl_3$/Ar plasma chemistry in an inductively coupled plasma/reactive ion etching (ICP-RIE) system at 30 W RIE power. An Agilent B1500 parameter analyzer was used to measure room temperature current-voltage characteristics on TLM (Transfer Length Method) pads of width 100 µm. Dielectric deposition was done at $250^0C$ using a Picosun SUNALE R-150B Atomic Layer Deposition tool. Low temperature I-V measurements were done using a Lake Shore cryogenic set-up equipped with liquid Helium closed-loop circulator.

Figure 2A is an image of as-grown $MoS_2$ film with mirror-like appearance due to its atomically smooth surface (rms roughness ~ 0.53 nm, 5 µm x 5 µm atomic force microscopy (AFM) scan, Figure 2B). In fact, the surface of as-grown $MoS_2$ sample reported here appeared to be smoother compared to that of 4-5nm $MoS_2$ film exfoliated from geological $MoS_2$.[38] The thickness of the $MoS_2$ film from sulfurizing 5nm Mo is approximately 7.0 nm based on AFM measurement (Figure 2C). The 2-θ/ω XRD scan (Figure 3A) exhibits only the (0001) family diffractions of $MoS_2$ and the diffraction of sapphire (0006) peak, which indicates a preferred growth orientation of $MoS_2$ with the c-axis parallel to that of sapphire substrate. Thickness fringes near the $MoS_2$ (0002) peak suggest a sharp interface (Figure 3A) and confirmed the thickness ~ 7 nm estimated from AFM scan (Supporting Information).

The off-axis (10-13) 2-θ/ω XRD scan (Figure 3B) across the full range of Ø=360⁰ shows six peaks at MoS$_2$ (10-13) position (Figure 3C) due to the six-fold symmetry of the hexagonal phase of MoS$_2$ and indicates the single-crystalline nature of as-grown MoS$_2$ film. A full range Ø=360⁰ scan of the sapphire (01-12) substrate taken with the sample in the same position showed three peaks of single-crystalline sapphire corresponding to its three-fold symmetry. The Ø-scans showed that the unit cell of MoS$_2$ was rotated by 30° with respect to that of sapphire substrate. To understand the in-plane orientation between MoS$_2$ and sapphire, we show the relative orientations of the two basal planes in Figure 4. The MoS$_2$ film and the sapphire substrate are rotationally commensurate: the length of 7 MoS$_2$ unit cells equals to the length of 8 sapphire unit cells after 30° rotation. The 30° rotation between the epi-layer and the substrate reduces the in-plane lattice mismatch to 13.0% with $\sqrt{3}a(MoS_2) = 5.47$ Å, a(sapphire) = 4.758Å. For the unrotated case, the lattice mismatch of MoS$_2$ and sapphire with a(MoS$_2$) = 3.16Å and a(sapphire) = 4.758Å would have been 50.5%. The lattice mismatch of 13% is still relatively high, but as shown previously,[39] Van der Waals epitaxial materials can tolerate a higher degree of lattice mismatch than that expected in traditional epitaxy.

Triple-axis rocking curve scan was also measured to confirm the single crystalline nature (Figure 3D). A full width at half maximum (FWHM) of 15.552 arc sec at MoS$_2$ (0002) diffraction was found to be about one-half of that of exfoliated single-crystalline MoS$_2$.[21] The narrow rocking curve FWHM suggests that the MoS$_2$ films grown by vapor-solid method had high crystalline quality with relatively low density of defects. The two characteristic Raman peaks of MoS$_2$ were observed with $E_{2g}^1$ at 381.2 cm$^{-1}$ and $A_{1g}$ at 406.5 cm$^{-1}$ with a peak separation of 25.3 cm$^{-1}$ confirming the film had the characteristics of bulk (or several-layer) MoS$_2$.[40]

The carrier mobility, which is strongly dependent on the crystalline quality as well as on the background impurity of the 2D film, plays a critical role in transport properties and hence on device performance. Theoretical calculations[41] and experiments[42, 43] show that while charge impurity limited scattering in single-layer MoS$_2$ is ~17 cm$^2$V$^{-1}$s$^{-1}$ without the high-κ dielectric screening effect, the intrinsic phonon-limited mobility of single-layer as well as multilayer MoS$_2$ is expected to be as high as 320-410 cm$^2$V$^{-1}$s$^{-1}$.[44, 45] Without the high-κ dielectric, few-layer MoS$_2$ films have been shown to have mobility from 10 to 100 cm$^2$V$^{-1}$s$^{-1}$ [46-48] which can be enhanced to several hundred cm$^2$V$^{-1}$s$^{-1}$ with a high-κ dielectric environment.[46, 47, 49] However, the high mobility values reported were on exfoliated samples, rather than on large area synthetic MoS$_2$. In this report, we show that high mobility approaching the phonon-limited values can be achieved using the synthesis method describe here.

In a log-log scale (Figure 5A), the current-voltage (I-V) curves exhibited two distinct slopes with a linear dependence on V at low-bias regime (0-10 V) and a quadratic dependence at higher (>20 V) bias regime. Further, for various TLM (Transfer Length Method) spacing (d), the current at higher bias was found to have $V^2/d^2$ dependence, indicating space-charge nature of the transport. The I-V curves were thus fitted with the space-charge transport equation:

$$I = \frac{q\,n\,\mu\,t\,L}{d} V + \frac{2\,\varepsilon_s\,\mu\,L}{\pi d^2} V^2 \qquad (3)$$

Here, q: electron charge, μ: electron mobility, t: thickness of MoS$_2$ film (= 7 nm), L: width of TLM pads (=100 μm), d: TLM pad spacing, $\varepsilon_s$: dielectric constant of bulk MoS$_2$ (=7.6).[48, 50] From the resulting fit (Figure 5B), an electron mobility of 120 (±20) cm$^2$/Vs and a carrier density of 2x10$^{11}$ cm$^{-2}$ were extracted. Interestingly, from the I-Vs measured in a direction perpendicular (in-plane) to the direction of measurement as reported above, an electron mobility of 65 cm$^2$/Vs and a carrier density of 2x10$^{11}$ cm$^{-2}$ were extracted by fitting I-Vs to equation (3).

This observation of anisotropic electron mobility (with a mobility-ratio of ~ 1.8) in few-layer MoS$_2$ in the two mutually perpendicular directions is in agreement with prior theoretical predictions[51] of anisotropic electron effective mass (0.53m$_0$ vs 0.73 m$_0$) for transport in ($\Lambda_{min}$) and in ($\perp\Lambda_{min}$) directions of the crystal.

To reduce the effect of surface-related defects and impurities on the mobility, the films were covered with 20 nm of Al$_2$O$_3$ by atomic layer deposition (ALD). The ALD Al$_2$O$_3$ can passivate some of the interface charge leading to less remote impurity scattering. The I-Vs measured at various temperatures between two device pads separated by 3.5 µm are shown in Figure 5C, and displayed typical characteristics of space charge transport. The extracted electron mobility showed weak temperature dependence while the carrier density was found to increase slightly from 1.3x10$^{11}$ cm$^{-2}$ at 10 K to 1.7x10$^{11}$ cm$^{-2}$ at 290 K. The room temperature electron mobility was found to be ~ 192 cm$^2$/Vs with ALD Al$_2$O$_3$ on MoS$_2$, an improvement from 120 cm$^2$/Vs which was extracted without dielectric on MoS$_2$.

A simple estimation of the electron mobility in few-layer MoS$_2$ was made based on polar optical phonon (POP) scattering,[52] acoustic deformation potential (ADP) scattering[53] and remote ionized impurity scattering in 2-dimensional electron gas,[53] assuming an electron effective mass of 0.53m$_0$. For remote impurity scattering, we assumed that the remote interface charge density (n$_{fix}^{2D}$) is located at the ALD/MoS$_2$ interface, and that the 2D carriers are in the middle of the few-layer film. Since the 2D sheet carrier in our MoS$_2$ film is non-degenerate (2x10$^{11}$ cm$^{-2}$), the carrier conduction does not take place predominantly at the Fermi Level. The scattering time was therefore estimated at any energy 'E' and weighed with density of states and Fermi-distribution over the entire energy range (details in Supporting Information).

Figure 5D shows the temperature dependent electron mobility calculated using POP, ADP and remote impurity scattering times compared with experimentally extracted mobility from our $MoS_2$ films. For a fixed remote impurity ($n_{fix}^{2D}$) of $1\times10^{11}$ cm$^{-2}$, the theoretically estimated mobility seems to have a close fit to those measured data points, validating our scattering time estimates. The $n_{fix}^{2D} = 1\times10^{11}$ cm$^{-2}$ which gives a good fit between theory and experiment, is also close to the actual carrier density ($1.3\times10^{11}$ -$1.7\times10^{11}$ cm$^{-2}$) extracted after ALD layer on $MoS_2$.

**Conclusion**

In conclusion, we reported the vapor-solid growth of high-quality few-layer $MoS_2$ films at $1100^0C$ using sulfur vapor obtained by decomposing $MoS_2$ powder. The as-grown surface was found to be atomically smooth and high resolution XRD scans and Raman spectroscopy indicated excellent out-of-plane ordering and epitaxial registry of the films over centimeter length scales. The film was found to exhibit space-charge transport and an electron mobility of 192 cm$^2$/Vs was extracted at room temperature. We also demonstrated anisotropic electron mobility (with a mobility ratio of ~ 1.8) in measurement directions mutually perpendicular to each other. ALD $Al_2O_3$ on $MoS_2$ was found to enhance the electron mobility which showed very weak temperature dependence from 10 K to 290 K. A simple scattering model based on optical phonon, acoustic phonon and remote impurity scattering was found to exhibit a good match with the experimentally extracted mobility. This demonstration of record high electron mobility for synthetic large area few-layer $MoS_2$ films is highly promising for enabling a wide variety of large-scale electronic device fabrication based on layered 2D materials.

## ACKNOWLEDGMENT

L.M. and Y.W. acknowledge the support from NSF (CAREER, DMR-0955471). D.N, E.L., C.H. Lee and S.R. acknowledge funding from the NSF NSEC (CANPD) Program (EEC0914790) and NSF Grant ECCS-0925529. Y.W. and S.R. acknowledge discussions with Professor James Speck (UC Santa Barbara) on the use of $MoS_2$ as a sulfur source.

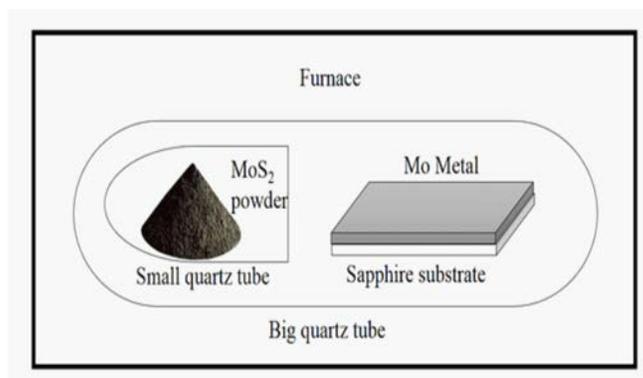

**Figure 1**. The schematic picture of growth set-up.

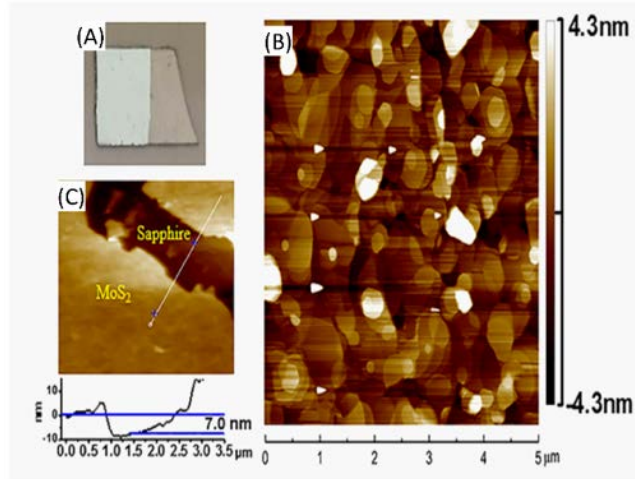

**Figure 2**. (A) Digital image of $MoS_2$ film; (B) AFM image of $MoS_2$ film; and (C) AFM images of $MoS_2$ near the edge to measure the thickness.

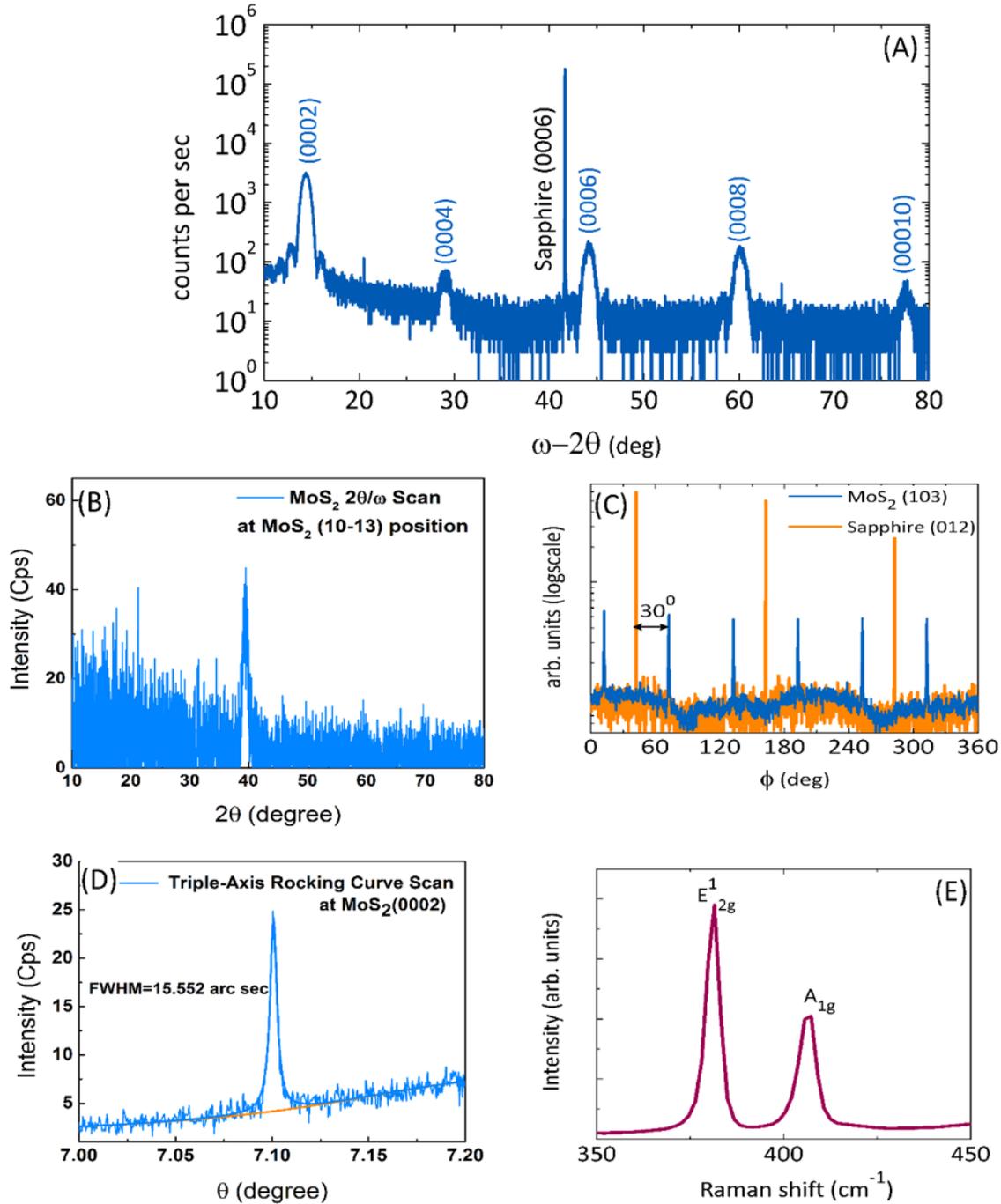

**Figure 3**. (A) 2Theta/Omega XRD Scan of MoS$_2$ film on sapphire (blue index: MoS$_2$ diffraction peaks, black index: sapphire (0006) diffraction peak.); (B) 2Theta/Omega XRD Scan at MoS$_2$ (10-13) position; (C) Phi Scan at MoS$_2$ (10-13) diffraction position and sapphire (01-12) position; (D)

Triple-Axis Rocking Curve Scan at MoS$_2$ (0002) diffraction position; and (E) Raman spectra of MoS$_2$ film.

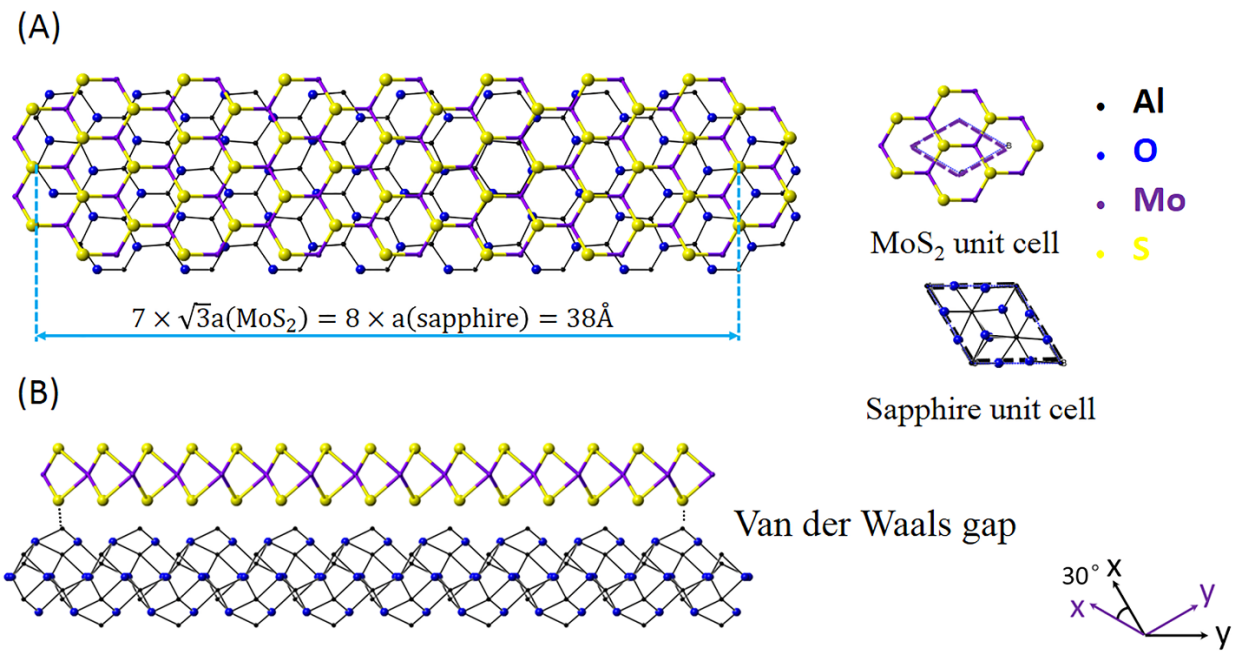

**Figure 4**. Relative in-plane orientation of MoS$_2$ to sapphire substrate with MoS$_2$ axis 30° with respect to sapphire: (A) top view and (B) side view. (Black-aluminum atom, blue-oxygen atom, purple-molybdenum atom, yellow-sulfur atom; purple dashed line-unit cell of MoS$_2$, black dashed line-unit cell of sapphire).

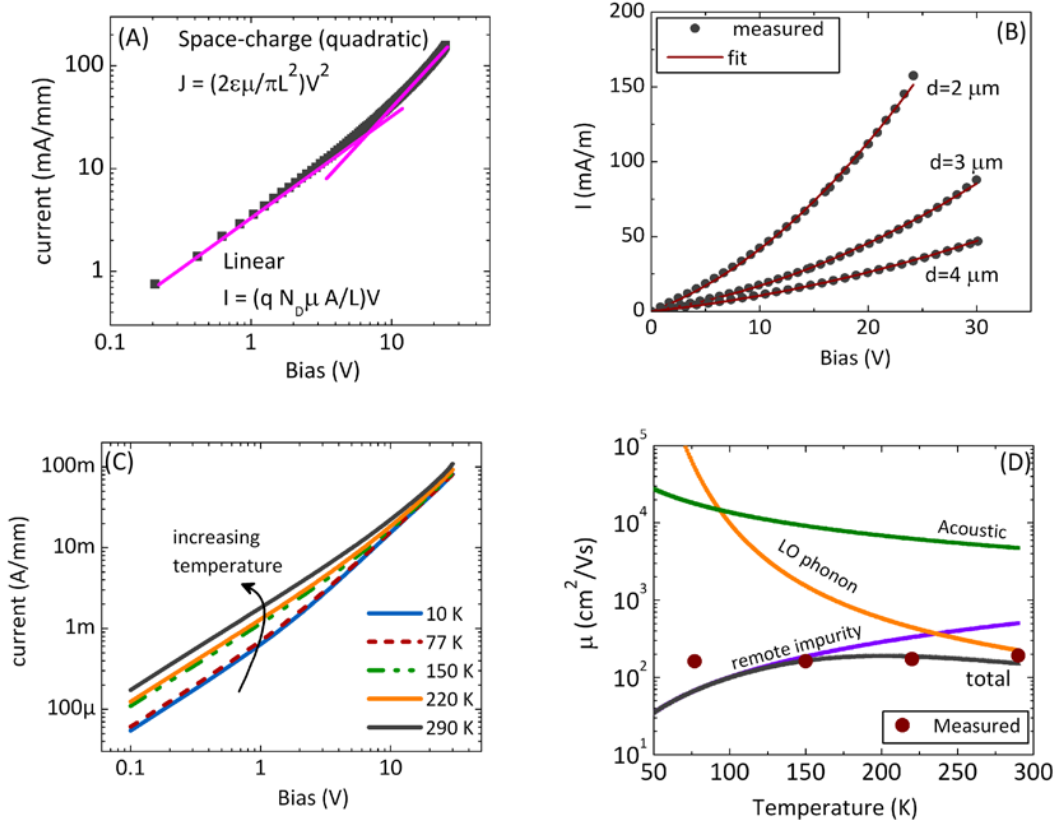

**Figure 5** (A) I-V log-log scale, showing two distinct slopes for linear and quadratic dependence on bias, (B) I-V measured at room temperature for various TLM spacing, along with the fit to I = BV+CV$^2$.; (C) Temperature dependent I-V for pad spacing of 3.5 µm; (D) Theoretical estimates of temperature dependent electron mobility limited by POP, ADP and remote impurity scattering compared with experimentally extracted mobility.